# Unexpected Near-Infrared to Visible Non-linear Optical Properties from 2-D Polar Metals.

*Megan A. Steves[1], Yuanxi Wang[2,3], Natalie Briggs[4,5,6], Tian Zhao[1], Hesham El-Sherif[7,8], Brian Bersch[4, 5, 6], Shruti Subramanian[4,5,6], Chengye Dong[5,6], Timothy Bowen[4,5,6], Ana De La Fuente Duran[4], Katharina Nisi[9], Margaux Lassaunière[10], Ursula Wurstbauer[10], Nabil Bassim[7], Jose J. Fonseca[11], Jeremy T. Robinson[11], Vincent Crespi[2,3], Joshua Robinson[3,4,5,6]\*, Kenneth L. Knappenberger, Jr.[1]\**

[1]Department of Chemistry, The Pennsylvania State University, University Park, PA 16802

[2]Department of Physics, The Pennsylvania State University, University Park, PA 16802

[3]2D Crystal Consortium, The Pennsylvania State University, University Park, PA 16802

[4]Department of Materials Science and Engineering, The Pennsylvania State University, University Park, PA 16802

[5]Materials Research Institute, The Pennsylvania State University, University Park, PA 16802

[6]Center for 2D and Layered Materials, The Pennsylvania State University, University Park, PA 16802

[7]Department of Materials Science and Engineering, McMaster University, Hamilton, ON L8S 4L8

[8]Mechanical Design and Production Engineering Department, Cairo University, Giza, 12613

[9]Walter Schottky Institute, Technical University of Munich, Am Coulombwall 4a, 85748 Garching

[10]Institute of Physics, University of Münster, Wilhelm-Klemmstr. 10, 48149 Münster

[11]U.S. Naval Research Laboratory, Washington, DC 20375

AUTHOR INFORMATION

**Corresponding Author** * Correspondence to: Kenneth L. Knappenberger, Jr. (klk260@psu.edu)

or Joshua Robinson (jar403@psu.edu)

ABSTRACT. Near-infrared-to-visible second harmonic generation from air-stable two-dimensional polar gallium and indium metals is described. The photonic properties of 2D metals - including the largest second-order susceptibilities reported for metals (approaching 10 nm$^2$/V) – are determined by the atomic-level structure and bonding of two-to-three-atom-thick crystalline films. The bond character evolved from covalent to metallic over a few atomic layers, changing the out-of-plane metal-metal bond distances by approximately ten percent (0.2 Å), resulting in symmetry breaking and an axial electrostatic dipole that mediated the large nonlinear response. Two different orientations of the crystalline metal atoms, corresponding to lateral displacements < 2 Å, persisted in separate micron-scale terraces to generate distinct harmonic polarizations. This strong atomic-level structure-property interplay suggests metal photonic properties can be controlled with atomic precision.

**TOC GRAPHICS**

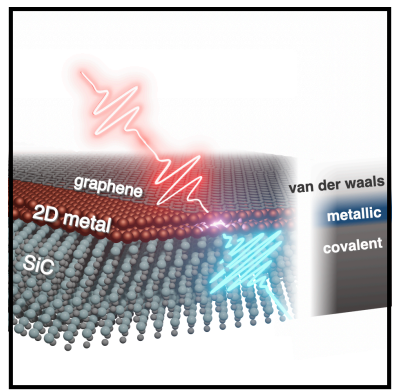

**KEYWORDS.** 2D materials, second harmonic generation, correlative microscopy, confinement heteroepitaxy, symmetry breaking.

TEXT. Nonlinear frequency mixing (e.g. harmonic generation) and polarization rotation of electromagnetic waves are the foundation of many important and emergent applications, including laser technologies, optical switches, chemical sensing, telecommunications, and frequency combs, among others [1]. The current state-of-the-art for second-order harmonic generation is achieved using a sequence of multiple quantum wells that are designed to enhance transitions resonantly at both fundamental and harmonic frequencies [2]. However, these systems are intrinsically limited to the mid infrared, precluding their operation at frequencies relevant for telecommunications, or where the resolution of molecular or biological imaging can be significantly enhanced. Therefore, strategies for achieving large nonlinear optical responses over a broader range of frequencies are needed. Although emerging two-dimensional semiconductors have shown great potential for nonlinear optical performance at higher frequencies, the reported second-order susceptibilities $\chi^{(2)}$ of these systems remain orders of magnitude smaller than those of planar multiple quantum wells [2–10].

Owing to their strong, broadband electromagnetic interactions, nanostructured metals offer the potential to establish a new state-of-the-art – in the visible and near-infrared – for nonlinear optical transduction. Achieving this new standard will require a predictive understanding of the interplay between material structure and nonlinear optical response. However, metal structure-property correlations outside of vacuum have been primarily limited to gold. Here, we report extremely efficient second harmonic generation from air-stable 2D metal heterostructures, with room-temperature near-infrared $\chi^{(2)}$ values approaching 10 nm/V. This extraordinary nonlinear response was obtained from a heterostructure consisting of crystalline, atomically thin group-IIIa metal films (gallium or indium) that are epitaxial to a SiC substrate and capped *in situ* with graphene.

The metal films extend over several hundred square microns adopting the hexagonal lattice of SiC, but the surprisingly large $\chi^{(2)}$-values are the result of subtle structural effects that occur on the atomic scale, which lead to axial symmetry breaking that is unique to the composite heterostructure—a hexagonal metal is not expected to yield large in-plane second order responses. This symmetry breaking, combined with quantum-confined electronic inter-band transitions resonant with the laser fundamental, results in the large in-plane $\chi^{(2)}$. The bonding character of the metal evolves from predominately covalent at the SiC/metal interface to primarily metallic between metal layers and non-bonded (van der Waals-like) at the metal/graphene interface, as reflected in an ~10% variation in the metal's layer-to-layer spacing (for three-layer gallium) along with an electrostatic gradient over two (indium) or three (gallium) atomic layers. This asymmetry leads to a theoretically predicted axial dipole and out-of-plane $\chi^{(2)}$ that is confirmed by angle-dependent second harmonic generation (SHG) measurements, verifying that these are 2D polar metals (2D-PMets). Density functional theory (DFT) calculations predict in-plane and out-of-plane second-order susceptibilities consistent with experimental measurements, as well as the presence of visible to near infrared electronic inter-band resonances, which may be leveraged to further amplify the intrinsically large nonlinear response of 2D-PMets.

In addition to efficient NLO transduction, polarization modulation is essential for efficient nonlinear optical switching. Correlative electron and polarization-dependent SHG microscopy resolves structural influences of the SiC substrate that are persistent into the metals and influence the far-field photonic properties of these heterostructures. Specifically, we show that Ångstrom-level defects, such as step edges, cause alternations in the metal film structure that extend for microns. As a result, spatially periodic recurrences of polarization-dependent NLO signals were observed. Taken together, these structural and optical analyses indicate that the efficient $\chi^{(2)}$ and polarization-dependent responses of the 2D polar metals are determined at the atomic level. Hence, control over interfacial bonding and structure may provide opportunities for further breakthroughs in room-temperature visible-to-near-infrared nonlinear optical technology that can be tailored at the atomic level using non-precious metals.

Two-dimensional Ga and In layers are formed through a process called Confinement Heteroepitaxy (CHet), where metal atoms intercalate to the interface between epitaxial graphene and SiC [11]. CHet forms crystalline, one-to-three-metal-atom-thick layers registered to the SiC. The native interface of epitaxial graphene on SiC contains dangling bonds which may be readily passivated through intercalation of metal atoms. The graphene layers not only aid in stabilizing the 2D metal films, but also protect them from oxidation, as confirmed in previous x-ray photoelectron spectroscopy studies [11]. Cross-sectional transmission electron microscopy (TEM) verifies that intercalation at the SiC/graphene interface yields crystalline metals 2–3 atomic layers thick, consistent with DFT phase stability calculations [11]. Using a 2D-Ga prototype (Figure 1A), charge densities calculated at the DFT level indicate that the first Ga layer bonds covalently with the SiC substrate; the bonding character then evolves to interlayer metallic bonding and finally van der Waals interactions (with weak charge transfer) between the top Ga layer and graphene [11]. This depth-dependent evolution in bonding character induces out-of-plane symmetry breaking throughout the metal layer, yielding a 2D polar metal (2D-PMet) with an intrinsic axial dipole. Importantly, the hermetic seal created by the graphene overlayer enables extensive ex-situ characterization of the 2D-PMet without degradation of the metal [11].

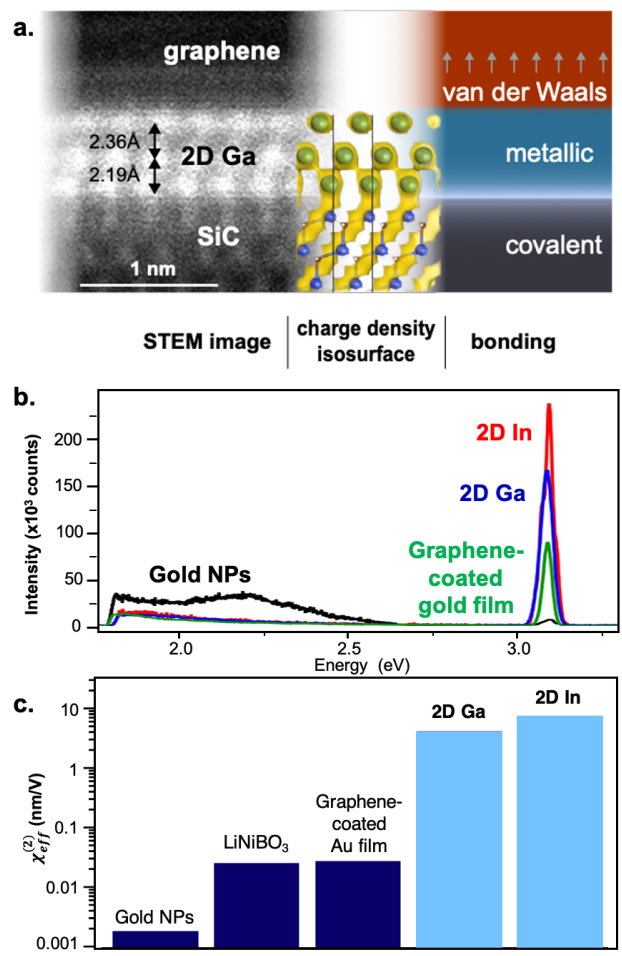

**Fig. 1.** Structure and nonlinear optical properties of 2D-PMets. (A) Cross-sectional STEM images reveal the layers of the SiC substrate/2D Ga/bilayer graphene structure (left). Upon intercalation, the Ga atoms adopt the hexagonal pattern of the SiC template, resulting in covalent substrate-metal bonding. The bonding character transitions to metallic between Ga layers, and to van der Waals at the Ga-graphene interface (right), resulting in an electrostatic gradient conveyed by the computed charge-density isosurface (center). This layer-by-layer evolution in bonding properties is accompanied by an approximate 10% change in metal lattice parameters, (2.19 Å to 2.36 Å lattice spacing), which is also captured by TEM (left). (B) Nonlinear spectra for 2D-PMets and 3D gold nanoparticles, demonstrating the enhanced SHG/TPPL ratio of 2D metals. (C) Comparison of $\chi^{(2)}$

values for 2D-PMets and other materials. All reported values are calculated using the average power of the fundamental.

Owing to the breaking of inversion symmetry by this out-of-plane atomic-level structure, 2D-PMets can function as strong second-order nonlinear optical transducers. This is evidenced by NLO spectra investigated through SHG imaging. Figure 1b compares NLO spectra for 2D-In and 2D- Ga to: (i) a 2.5-nm-thick Au film on SiC encapsulated with CVD graphene (Methods) and (ii) 3D gold nanorods, which are standard environmentally stable metallic nanostructures for investigating NLO response [12]. 2D-PMets are characterized by prominent NLO peaks at 3.1 eV, resulting from second harmonic generation of the 1.55 eV fundamental, which was confirmed from quadratic signal power dependences (Figure S1A). In contrast, the NLO spectra of three-dimensional gold nanorods are dominated by broad two-photon photoluminescence typical of colloidal coinage metals [13]; SHG is a minor contribution to the global NLO intensity. The integrated intensities of the experimentally measured SHG and a polarizable sheet model (Methods) were used to calculate the magnitudes of the per-volume second-order susceptibility $\chi^{(2)}$ for these systems, which are compared in Figure 1C [14] For 2D-PMets, values of 4.6 and 8.5 nm/V are obtained for 2D-Ga and 2D-In, respectively. These values are calculated using metal thicknesses of 0.60 and 0.45 nm, as estimated from cross-sectional TEM [11]. While SiC and epitaxial graphene also exhibit nonlinear signal, the SHG yield from graphene/SiC is small compared to that from a 2D-PMet-based heterostructure. Both 2D-PMets show $\chi^{(2)}$ more than 1000× those of 3D gold nanorods, and over 100× larger than graphene-coated Au films and industrial standards such as $LiNbO_3$ [14]. Because $\chi^{(2)}$ values are intrinsically higher for lower bandgap materials [15], we also compare the 2D-PMets to other metal-surface SHG sources using per-area units (nm²/V) [16,17]. The in-plane components of 2D-Ga and 2D-In are $\chi^{(2)}_{yyy}$ = 4.8 and 3.8 nm²/V, respectively, which is reasonably consistent with DFT ($\chi^{(2)}_{yyy}$ = 3 nm²/V for 2D In, see Methods and Supplemental information). Furthermore, 2D-PMets are predicted to be more efficient than other metal-surface SHG sources like Al(111), as the theoretical estimate of $\chi^{(2)}_{yyy}$ for the Al(111) surface is 0.2 nm²/V [18]. Hence, the Ga and In 2D-PMets exhibit the largest $\chi^{(2)}$ reported for any metal transducer.

With the extra-ordinarily efficient nonlinear optical response of 2D-PMets established, we next confirm the importance of their out-of-plane symmetry breaking in producing the surprisingly large $\chi^{(2)}$ values. Since centrosymmetric crystals have zero $\chi^{(2)}$, bulk Ga and In with $D_{2h}$ (mmm) and $D_{4h}$ (4/mmm) point group symmetries (crystal classes) respectively are not expected to generate SHG. Even considering a hypothetical freestanding 2D metal with hexagonal packing as imposed by the SiC substrate, two and three atomic layers with AB and ABC stacking belong to the $C_{6h}$ point group, which is centrosymmetric and thus not SHG-active. The actual inversion symmetry breaking of the 2D-PMets relies on the out-of-plane polarity, as verified using the angular dependence of the p- and s-polarized SHG intensities ($I_p/I_s$). The electric field of the fundamental is aligned with the axial dipole by internal reflection at the sample plane using the objective by beam translation parallel to the polarization vector (Figure 2A). If an axial dipole is present, the SHG ratio $I_p(2\omega)/I_s(2\omega)$ is expected to increase quadratically with increasing beam displacement from the center of the objective back aperture [19]. Figure 2B, which portrays the angular dependence of one of the striped regions from Figure 2C, demonstrates this to be the case (the origin of this stripe pattern is discussed below). In contrast, no angular dependence was detected for 5 nm thick gold films encapsulated with graphene; its out-of-plane $\chi^{(2)}_{zzz}$=0.1 nm²/V [16] is similar in magnitude to its in-plane $\chi^{(2)}_{yyy}$=0.1 nm²/V, and hence, no angular dependence is observed. The axial dipole revealed by the Figure 2B data is thus a distinct feature of the CHet-formed 2D-PMets, one that is likely associated with the strong layer-by-layer gradient in bonding character, and results in the formation of efficient second-order nonlinear optical transduction.

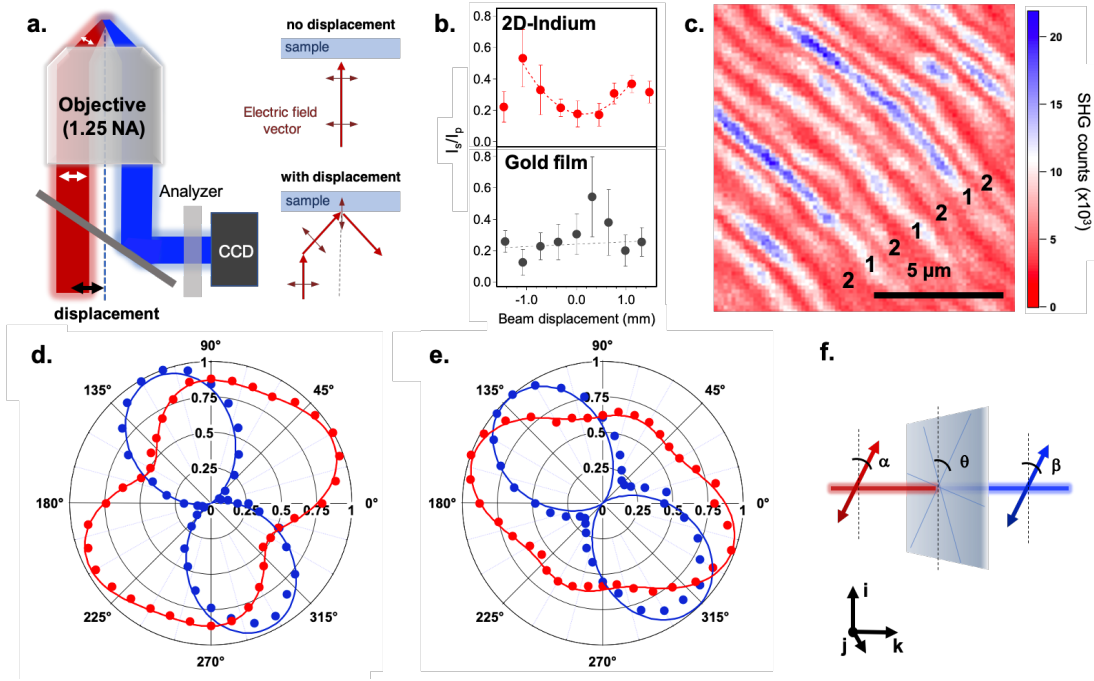

**Fig. 2.** Probing the structure of 2D-In using angle- and polarization- resolved SHG microscopy. (A) Experimental setup for angle-resolved SHG measurements, used to obtain (B) ratio of $I_p/I_s$ for one type of stripes as a function of beam displacement. The dashed line is a guide to the eye illustrating the approximate quadratic dependence of the data. The ratio of $I_p/I_s$ for a gold film is shown for comparison. (C) Example SHG-detected image of 2D-In. Distinct regions, labelled 1 and 2, demonstrate different polarization responses. (D, E) Representative examples of excitation (red) and emission (blue) polarization-dependent SHG intensities obtained by changing $\alpha$ and $\beta$, respectively for Type 1 (D) and Type 2 (E) regions. Filled circles show the experimental data, while the dashed lines are fits to the data using the parameters in Table S1. (C) Schematic of polarization-resolved experimental setup, depicting the angles used in the polarization analysis. The angle of the crystal axis with respect to the laboratory $\hat{\imath}$ axis is given as $\theta$. $\alpha$ and $\beta$ are the polarization of the fundamental and harmonic waves, respectively, within the sample plane and referenced to the $\hat{\imath}$ axis.

The intrinsically large second-order response of the polar metals is further increased by electronic states that resonate at the fundamental frequency. These states are localized within two atomic layers of the metal/SiC interface. This is shown by a detailed band decomposition of $\chi^{(2)}_{yyy}$ using 2D-In as a prototypical 2D PMet, where bulk and surface contributions to $\chi^{(2)}_{yyy}$ were found to be insignificant for thicker metal layers on SiC (see Supplemental information). For completeness, the out-of-plane component is estimated from theory and compared to other metal surfaces in Supplemental information. Owing to their atomic-level geometric and electronic structure, the $\chi^{(2)}$ of 2D- PMets are, to our knowledge, the largest reported for metallic transducers, and may provide a platform for achieving near-infrared second-order responses that can compete favorably with the impressive mid-infrared values obtained for multiple-quantum-well structures [2,9].

The substrate is a dominant factor in determining the interfacial structure, and in turn, the optical response of the 2D-PMet. Beyond the large $\chi^{(2)}$ values, this influence is responsible for the formation of polarization-dependent SHG signals. This behavior is portrayed in Figure 2C, where two types of regions with distinct polarization patterns are detected in 2D-In. The two regions, which have similar maximum SHG intensities, alternate spatially in striped patterns (labeled 1 and 2), with distinct and complementary SHG-intensity polarization dependences (Figures 2D,E and Movies S1 and S2). Analysis of these polarization patterns in terms of a rank-three nonlinear susceptibility tensor (Methods) provides evidence for the origin of the striping. The polarization and crystal axes are defined with respect to the lab frame in Figure 2F, θ being the angle of the 2D-PMet crystal axis with respect to the sample plane and α and β the polarization angles of the fundamental and harmonic within the sample plane. Fitting the SHG excitation and emission polarization patterns (Figs. 2D,E) retrieves both θ and the relative magnitudes of each element of the nonlinear susceptibility tensor for each stripe type (Methods). Compared to the expectations for the major crystal classes, the $\chi^{(2)}$ elements obtained by fitting (Table S1) most closely match a polar $C_{3v}$ (3m) point group symmetry for both types, but with moderate deviations resulting from defects. This correspondence with $C_{3v}$ structure provides further evidence that the 2D-In is polar and adopts the in-plane symmetry of the SiC substrate. Similar polarization dependences are observed in 2D-Ga (Figure S1B, C), which implies that substrate templating of the metal is a general phenomenon.

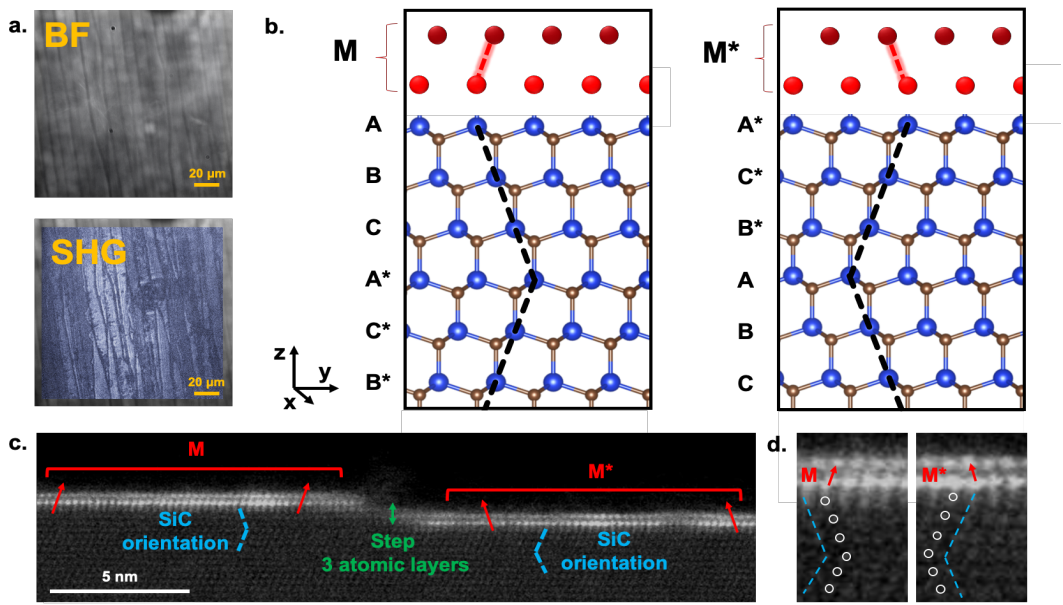

**Fig. 3.** Structural origin of the stripes observed in polarization-resolved SHG. (a) correlative brightfield (top) and SHG (bottom) imaging demonstrating that step edges separate different regions. (b) schematic depicting the In (red) orientations M and M*, where the first layer of In is epitaxial to the silicon (blue) and the second layer is registered over the uppermost carbon (brown) atoms. (c) STEM image of 2D-In cross-section, demonstrating the change from M to M*

orientation at a SiC step edge. (d) Larger view of two areas selected from panel c focusing on the M and M* orientations. Circles indicating the approximate location of Si atoms and dashed lines are added as a guide to the eye.

Correlated bright-field and SHG microscopy demonstrate that the stripe regions are separated by step edges in the SiC substrate (Figure 3A). Comparison of optical and atomic force microscopy images also support this interpretation, indicating the observed SHG stripes result from metal terraces that persist for several microns (Figure S2). Cross-sectional scanning transmission electron microscopy (STEM) analysis confirms that the sample in Figure 2A is predominantly two atomic layers of In (Figure S3). While some areas with other thicknesses are observed, these are not correlated with SiC steps, and hence, we rule out differences in metal thickness as the cause of the alternating polarization patterns. On the other hand, 6H-SiC is known to alternate its layer stacking order every three SiC units (Figure 3B). Taken together with the registry of the first- and second-layer metal atoms above the uppermost Si and C atoms, respectively (Figures 1 and S3), we infer that the structural alternation in the SiC substrate results in a metal lattice rotation of 180º at every three-unit-high step in the SiC substrate. The result is two possible metal orientations relative to the first layer of carbon in the SiC substrate, which we label M and M*, which have the in-plane nonlinear susceptibility elements $\begin{pmatrix} & & xxy \\ yxx & yyy & \end{pmatrix} = \begin{pmatrix} 1 & -1+n \\ -1+n & 1 \end{pmatrix}_M$ and $\begin{pmatrix} & & 1+n \\ 1+n & -1 & \end{pmatrix}_{M*}$, where $n$ represents deviations from $C_{3v}$ symmetry caused by defects (for example, 1D-step edges along $(1\bar{1}0n)$ SiC). Inspection of the $\chi^{(2)}$ elements obtained from fitting of the polarization patterns of 2D-PMets for type 1 and 2 stripes reveals them to be consistent with this model and allows us to observe that the contribution from defects is consistent with $C_{1h}$ symmetry. Thus, we conclude that alternating metal orientations combined with in-plane symmetry-breaking defects cause the distinguishable striped regions observed with polarization-resolved SHG microscopy. We also directly observe the inversion of metal orientation at a step edge through STEM imaging of 2D-In (Figure 3C). Taken together, correlative polarization-resolved and angle-dependent SHG and electron microscopy, along with theoretical calculations, implicate unique roles for out-of-plane and in-plane symmetry breaking for determining the optical responses of the 2D-PMets: whereas out-of-plane structure is fundamental to the large SHG signals, in-plane symmetry breaking determines their distinct polarization response. Although the interfacial structure determined the SHG polarization response, the $\chi^{(2)}$ values were uniform across the 2D-PMet. This observation implies that synthetic control over the lattice properties could lead to efficient polarization-selective nonlinear optical properties of 2D metals.

The results presented here indicate that 2-D polar metals formed by confinement heteroepitaxy could provide a materials platform for unprecedented nonlinear optical transduction at the technologically important visible to near-infrared frequency range. Moreover, we find that the surprising nonlinear optical properties of these systems result from their unique bonding and lattice properties, which are determined with structural precision on the atomic level. These Ångstrom-scale effects extend uniformly through the metals on a micron length scale, providing rich second-order nonlinear optical properties. Because the SiC-metal interaction determines both the in-plane and out-of-plane symmetry of 2D-PMets, there is strong potential for tuning material nonlinear susceptibilities (e.g. $\chi^{(2)}$) by controlling substrate-metal interactions, and the air stability of the CHet platform (through graphene encapsulation) means that many metals beyond Au are now candidates for study, optimization, and utilization of NLO response. The SHG polarization and angular dependences observed here are unique to the atomic-level interfacial structure of the 2D polar metals. Hence, 2D-PMets may have intrinsic advantages for optical switching applications and for forming nonlinear metasurfaces. In particular, defects were shown to induce lateral metal atom displacements of < 2 Å that persisted in terraces for several microns. These subtle structural differences resulted in the generation of harmonic signals with distinct polarizations. Therefore, structural changes at the atomic level may serve as a template for controlling the lattice and photonic properties of metals. NLO transduction by these metals could be further enhanced by in-plane resonant electronic excitation such as visible-to-near infrared inter-band excitations (Figure S4) and plasmon excitations [20]. For the latter, the SHG response can also be increased by local surface fields. These enhancements occur when the real part of the metal dielectric function ($\varepsilon_1$) is negative [21,22]. In addition, for sharp plasmon resonances, inter-band contribution to the imaginary part ($\varepsilon_2$) must be small so that plasmon dephasing times are long (i.e. low damping from inter-band transitions). Importantly, we find the metal quantum confinement yields low $\varepsilon_2$ and oscillatory $|\varepsilon_1|$ at discrete optical or near-IR excitation energies for 2D-PMets (Figure S5). Moreover, calculated $\varepsilon_1$ values take on large negative values for excitation spanning 0.5 eV to 1.5 eV, indicating that quantum confinement may intensify 2D-PMet resonances at specific vis-to-NIR energies. This expectation is supported by the theoretically predicted plasmon response for 2D-PMets (Figure S5). Thus, is it highly likely that further amplification of material nonlinear responses can be achieved through resonance matching of the plasmon with the fundamental ($\omega$) or harmonic ($2\omega$) frequencies. Generation of an enhanced electric field by excitation of the plasmon may also be possible. Moreover, the atomic thickness of 2D-PMets leads to quantized resonances that can be used to form hybrid resonances in multi-component heterojunctions. These quantized hybrid states could be leveraged to mediate efficient nonlinear optical processes at wavelengths important to telecommunications and optoelectronic applications.

**Experimental Methods**

Synthesis of 2D Metals

2D-Ga and In layers were intercalated to the epitaxial graphene/silicon carbide interface through a thermal vaporization-based method [11]. Epitaxial graphene/silicon carbide was first synthesized through a high-temperature Si sublimation process. The resulting graphene was then treated with an $O_2$/He plasma, which induces defects in the graphene layers. Following this treatment, 30–60 mg of Ga or In metal was placed in the base of a custom-made alumina crucible designed to hold 1×1 cm epitaxial graphene/silicon carbide substrates. The defective graphene layers were placed in the crucible, with the graphene layers facing downward over the Ga/In metal. The crucible was then loaded into STF-1200 tube furnace equipped with a 1″ O.D. quartz tube. The furnace was evacuated to ~5 mTorr and subsequently pressurized to 300 Torr with Ar. The furnace was then heated to 600–800°C and held for 30 minutes under 50 sccm of continuous Ar flow. The furnace was then cooled to room temperature and the intercalated epitaxial graphene/silicon carbide samples were removed.

Encapsulated gold films

To compare 2D polar metals with conventionally deposited metal films, we electron-beam evaporated Au films (2.5–20nm thick, 1Å/s) onto SiC substrates and then capped the Au film with a transferred CVD graphene film. These samples were then annealed in a tube furnace in flowing Ar/$H_2$ gas between 250–300°C to induce a recrystallization of the encapsulated Au film [23]. This process results in the microstructure of the Au film transforming from nano- to micro-crystalline and the Au layer crystallographically aligning with the capping graphene layer [23]. The CVD graphene film was grown on Cu foil using $CH_4$/$H_2$ gas with ~50mtorr pressure at 1030°C.

Cross-section preparation

Specific sites with high-intensity signal from the SHG map were located in the SEM with the help of fiducial metal particles on the surface that were present in the optical images acquired before/during the SHG map. Cross-sections from the specific sites are prepared by a Helios G4 PFIB UXe DualBeam with a Xe+ plasma source for the ion column to avoid contamination from typical Ga FIB. A ~50 nm platinum coating is deposited by the electron beam deposition at 5 KeV and 6.4 nA. Then, a 3 μm Tungsten is deposited by the ion beam at 16 KeV followed by and 1-μm platinum/carbon layer at 30KeV. Subsequently, the cross-sections were prepared using the standard preparation steps.

STEM

High-resolution STEM of the prepared cross-sections was performed in a FEI Titan 80–300 Cubed transmission electron microscope equipped with spherical aberration correctors in both the probe and the image forming lenses in addition to a high-brightness gun (XFEG). Operational conditions are optimized to minimize the electron beam damage at the interface for imaging. STEM images are collected at 300KeV and a dose rate of less than 60 e/ Å$^2$/ s using an in-column Fischione HAADF detector (model 3000) at 19.1 beam convergence angle, 50.5–200 mrad collection angles at 115 mm Camera length, and 50 μm C2 aperture.

Theory estimate of in-plane $\chi^{(2)}$

In-plane $\chi^{(2)}$ for metal surfaces are rarely modeled or measured since they typically vanish for isotropic metallic surfaces such as (001) facets or disordered surfaces. A finite in-plane $\chi^{(2)}$ due to an inversion breaking surface (e.g. (111) surfaces) has no Drude-like contributions since, by construction, free-electron states is agnostic to the presence of lattice effects. In-plane $\chi^{(2)}$ are therefore mainly due to interband transitions and are best treated by the standard sum-over-states formalism developed by Sipe and co-workers [24–26]. All DFT-level calculations were performed using the ABINIT package [27,28]. The ground-state properties and response functions were calculated within the generalized gradient approximation using norm-conserving pseudopotentials with an energy cut-off of 952 eV and a k-point mesh of 24×24×1. The optical spectra achieved convergence with 210 total bands and a smearing of 80 meV. At 1.5 eV, the calculated $\chi^{(2)}_{yyy}$ is 3 nm$^2$/V.

Optical Measurements

Optical measurements were performed using a homebuilt nonlinear optical microscope previously described [12,13,29]. All nonlinear optical measurements use a 800 nm fundamental generated by a mode-locked 80MHz Ti:Sapphire laser (Tsunami, Spectra-Physics). Imaging and spectral analysis of optical signals was performed using a coupled spectrometer/electron multiplying charge-coupled device (Shamrock 303i/ iXion Ultra 897, Andor Technology). The power of the 800 nm fundamental was adjusted using a neutral density filter. The fundamental was focused onto the sample with a 0.23 NA aspheric lens, and the resulting signal was collected with a 1.25 NA oil-immersed objective. A 680nm short pass filter was used to exclude the fundamental photons before the detector. The nonlinear spectra were measured at different excitation powers (40mW to 100mW). The integrated intensity of the peak at 3.1 eV in the nonlinear spectra had a quadratic dependence on power (Figure S1), confirming that it resulted from SHG.

For SHG imaging, a 3.1eV bandpass filter was placed before the detector to isolate SHG photons. The incident power was set at 80 mW, and the image was acquired with 10 accumulations at 0.1 s exposure time, with 200 EM gain.

Polarization analysis was performed for a 2D-In sample. For the polarization dependence, a linear polarizer and half-wave plate were placed before the sample. The incident polarization was adjusted in 10° increments with an image collected at each step (Movie S1). For the emission polarization dependence, the incident polarization was fixed at 180° with respect to the lab frame and an analyzer was placed before the detector. The analyzer was rotated in 10° increments and an image collected at each step (Movie S2). To generate the polar plots of Figures 3a and b, a stripe from the SHG image was selected as an ROI and the average SHG intensity in that ROI at each step of the half-wave plate or polarizer was determined. Polar plots were generated for five stripes of each type, resulting in 10 polar plots (five excitation polarization and five emission polarization) of each type of stripe for use in fitting.

Incident-angle polarization analysis was performed in an epi-geometry (shown in Figure 2b). A linear polarizer and half wave plate were used to polarize the 800 nm fundamental along i direction of the laboratory axis. A mirror mounted on a translation stage was used to displace the beam along the i direction of the laboratory frame. The sample was rotated about the i axis so that the incident polarization was roughly 45° from the crystal x axis, ensuring $I_x(\omega) = I_y(\omega)$ as $I_z(\omega)$ increases. An analyzer was placed before the detector, and a series of images was collected by rotating the analyzer in 10° increments at each beam displacement. Image series were collected at 9 total displacements, from −1.45 to +1.46 mm. Based on an estimate of the beam diameter as ~1.5mm, a displacement of the beam of 0.8 mm is expected to result in total internal reflection of the fundamental wave, and in-plane excitation of the sample. Displacement distances smaller than 0.8 mm result in gradual in-plane angular displacements of the fundamental. Displacement distances exceeding 0.8 mm do not result in internal reflection. This explains the deviation from the expected quadratic dependence of $I_p/I_s$ at −1.45 and +1.46 mm displacement.

Calculation of $\chi^{(2)}$

To calculate $\chi^{(2)}$, the signal in counts, was used to calculate the total energy of the harmonic photons detected over the collection time according to the following equation:

$$E_v = \left(\frac{cts}{g}\right) \times \left(\frac{S}{QE}\right) \times (3.65) \qquad (1)$$

where $E_V$ is the total energy of the photons incident on the detector, cts is the number of counts (per pixel) detected by the EMCCD, g is the amount of electron multiplying (EM) gain applied, S is the CCD sensitivity, QE is the quantum efficiency of the detector in the spectral region of interest, and 3.65 is a physical constant for electron creation in silicon [30,31]. In this procedure, g was set at 200, S equaled to 4 and the quantum efficiency, QE, was 71% at 400nm. The number of counts was retrieved from the measured SHG image. The energy in eV detected during the 1s total collection time was converted into the power of the second harmonic signal $P_{2\omega}$.

To estimate the second-order nonlinear susceptibility of 2D PMets, we use the sheet model. Treating the 2D metal as a polarizable sheet, the $\chi^{(2)}_{sheet}$ is given by: [10,22]

$$\left|\chi^{(2)}_{sheet}\right|^2 = \frac{P_{2\omega}\left[n^{SiC}_{\omega}+1\right]^6 \epsilon_0 c^3 A}{2048\pi^3 \omega^2 P_{\omega}^2} \qquad (2)$$

Where $\omega$ is the fundamental frequency, $P_{2\omega}$ is the measured SHG power, $P_{\omega}$ is the incident laser power at the fundamental frequency, $n^{SiC}_{\omega}$ is the refractive index of the SiC substrate at the fundamental frequency, and $A$ is the area illuminated by the exciting pulse. The nonlinear susceptibility of the sheet is then converted into an effective nonlinear susceptibility for the material, $\chi^{(2)}_{eff}$, by $\chi^{(2)}_{eff} = \frac{\chi^{(2)}_{sheet}}{d}$, where $d$ is the thickness of the nonlinear material. Refractive indices $n$ of the materials at various frequencies were obtained from the literature [32,33].

Polarization analysis

Analysis of the polarization patterns was performed by considering the 27-element rank 3 nonlinear susceptibility tensor. The angles of the incident polarization and crystal axes with respect to the laboratory frame referenced in the following are defined in Figure 3c. By applying the intrinsic permutation symmetry for SHG (i.e., $\chi^{(2)}_{ijk}=\chi^{(2)}_{ikj}$), the $\chi^{(2)}$ tensor is reduced to 18 elements.

Due to the spectral overlap of our fundamental with the plasmon resonance of the 2D metals, Kleinman symmetry cannot be reliably applied. We further assume that the crystal axis z is aligned with the k axis of the laboratory frame, and that dispersion of the linearly polarized incident light by focusing with a low NA aspheric lens is negligible. Therefore, the electric field $E_z$ is minimal, and the nonlinear polarization $P^{(2)}$ induced by linearly polarized light with angle α with respect to the laboratory frame is given by

$$P^{(2)} = \epsilon_0 \begin{pmatrix} \chi_{xxx}^{(2)} & \chi_{xyy}^{(2)} & \chi_{xzz}^{(2)} & \chi_{xyz}^{(2)} & \chi_{xxz}^{(2)} & \chi_{xxy}^{(2)} \\ \chi_{yxx}^{(2)} & \chi_{yyy}^{(2)} & \chi_{yzz}^{(2)} & \chi_{yyz}^{(2)} & \chi_{yxz}^{(2)} & \chi_{yxy}^{(2)} \\ \chi_{zxx}^{(2)} & \chi_{zyy}^{(2)} & \chi_{zzz}^{(2)} & \chi_{zyz}^{(2)} & \chi_{zxz}^{(2)} & \chi_{zxy}^{(2)} \end{pmatrix} \begin{pmatrix} E^2 \cdot \cos^2(\theta - \alpha) \\ E^2 \cdot \sin^2(\theta - \alpha) \\ 0 \\ 0 \\ 0 \\ 2E^2 \cdot \cos(\theta - \alpha)\sin(\theta - \alpha) \end{pmatrix} \quad (4)$$

where θ is the angle of the crystal x-axis with respect to the laboratory frame [34]. The intensity of the second harmonic polarized with respect to the x, y, and z crystal axes is given by

$$I_x^{SHG} \propto [\chi_{xxx}^{(2)} \cos^2(\theta - \alpha) + \chi_{xyy}^{(2)} \sin^2(\theta - \alpha) + 2\chi_{xxy}^{(2)} \sin(\theta - \alpha)\cos(\theta - \alpha)]^2 \quad (5)$$

$$I_y^{SHG} \propto [\chi_{yxx}^{(2)} \cos^2(\theta - \alpha) + \chi_{yyy}^{(2)} \sin^2(\theta - \alpha) + 2\chi_{yxy}^{(2)} \sin(\theta - \alpha)\cos(\theta - \alpha)]^2 \quad (6)$$

$$I_z^{SHG} \propto [\chi_{zxx}^{(2)} \cos^2(\theta - \alpha) + \chi_{zyy}^{(2)} \sin^2(\theta - \alpha) + 2\chi_{zxy}^{(2)} \sin(\theta - \alpha)\cos(\theta - \alpha)]^2 \quad (7)$$

The contribution of $I_z^{SHG}$ was considered to be negligible in our experimental geometry. The SHG intensity as a function of β is given by:

$$I(\beta)_{total}^{SHG} \propto \left[\cos(\beta - \beta_0) \cdot P_x^{(2)} + \sin(\beta - \beta_0) \cdot P_y^{(2)}\right]^2 \quad (8)$$

where $\beta_0$ is a reference angle between the analyzer and crystal axis. Analysis of the normalized polar plots for type-1 and type-2 stripes was performed separately. The polar plots were fit with the constraints that each $\chi^{(2)}$ element and θ be equal for the excitation and emission plots of identical stripes. $\chi^{(2)}_{yyy}$ was set to 1. The resulting fitting parameters and standard deviations obtained from fitting 5 examples of each type of stripe are summarized in Table S1.

Atomic Force Microscopy

Atomic force microscopy was performed using a Bruker Icon AFM at a scan rate of 0.452 Hz, peak force set point of 1.497 nN, and peak force amplitude of 150 nm.

Experimental determination of 2D metal dielectric function

The dielectric functions of 2D-In and 2D-Ga were determined by spectroscopic ellipsometry (SE) using a M-2000 ellipsometer (J.A. Woollam) in RCE-mode. The measurements were implemented at wavelengths ranging from 300 to 1700 nm with a spectral resolution of 5 nm, with an incident angle of 55° with a macroscopic spot size (~ 1mm). Data was collected under ambient conditions in a clean room. The obtained data was modeled using the EP4 modelling program from Accurion GmbH. The dielectric properties of the 2D metal films are described by Lorentzian and Drude functions.

ASSOCIATED CONTENT

**Supporting Information**.

The following files are available free of charge.

Materials and Methods (PDF)

Supplementary discussion and figures (PDF)

Supplementary Movies 1 and 2 (.avi)

AUTHOR INFORMATION

**Notes**

The authors declare no competing financial interests.

ACKNOWLEDGMENT

This research was supported by the Air Force Office of Scientific Research, grant number FA-9550-18-1-0347 and the National Science Foundation Graduate Research Fellowship Program under Grant No. DGE1255832. This work was also supported by National Science Foundation, award number CHE-1807999. Funding for this work was also provided by the Northrop Grumman Mission Systems' University Research Program, Semiconductor Research Corporation Intel/Global Research Collaboration Fellowship Program, task 2741.001, National Science Foundation (NSF) CAREER Awards 1453924 and 1847811, the Chinese Scholarship Council, an Alfred P. Sloan Research Fellowship, NSF DMR-1708972, DMR-1420620, and 1808900, and the 2D Crystal Consortium NSF Materials Innovation Platform under cooperative agreement DMR-1539916. Funding for the McMaster work was by AFOSR Award FA9550-19-1-0239 and the NSERC - Natural Sciences and Engineering Research Council of Canada Discovery Grant program. All electron microscopy was performed at the Canadian Centre for Electron Microscopy. The work at NRL was funded through the Office of Naval Research.

## Supplementary Text

Theory determination of $\chi^{(2)}_{yyy}$ and its spatial origin

DFT-level interband contributions to $\chi^{(2)}_{yyy}$ for 2D indium are shown in Figure S6. The imaginary part of $\chi^{(2)}$ is plotted because of its potential usefulness when identifying interband resonances (*35*), and the absolute value is plotted to compare with the experimental measurement of $\chi^{(2)}$. The per-area $\chi^{(2)}_{yyy}$ =3 nm²/V for two-atomic-layer thickness (2L) of indium is estimated from $\chi^{(2)}$ values within the energy range covered by the gray box 1.5±0.2 eV.

To demonstrate that the high $\chi^{(2)}$ values of 2D PMets rely on the metal/SiC interface, we compare $\chi^{(2)}$ of 2L-In with thicker layers of In (5,9,15 layers), each adopting the same hexagonal lattice as 2L In and relaxed within DFT. The observation that $\chi^{(2)}$ of these larger systems does not scale with thickness rules out the possibility that the large $\chi^{(2)}$ was dominated by bulk effects, leaving the possible origins of large $\chi^{(2)}$ to be either the metal/SiC interface or the metal surface.

In order to isolate the interface contributions to the large $\chi^{(2)}$ for the 2D metals, the $\chi^{(2)}$ components for a series of 2L Indium thicknesses were decomposed in reciprocal space and by electronic band, as described in the following. Figure S7a shows the reciprocal-space-decomposed $\chi^{(2)}$ for 15L-In/SiC within ω=1.5±0.2eV, where the dominant contributions appear not at K, but close to K. Further breaking down $\chi^{(2)}$ by band contribution is shown in Supplemental Figure S7b, where only the dominant transitions at 2ω are visible, and where red and blue indicate contributions of opposite sign. The dominant contribution near K mentioned above can now be found on a pair of bands along Γ–K. Comparison with the atomic-orbital-projected band structure in Figure S7c revealed that the initial state of this dominating transition is indeed localized at interfacial In and Si atoms. Other strong transitions from lower-energy bulk bands at exactly K can also be seen, but these do not add constructively to a significant summed contribution at K.

Similar trends are found for 13L- and 9L-In/SiC: near-K interband transitions originating from interfacial In and Si atoms have a consistently dominant and positive contribution to $\chi^{(2)}$ at ω=1.5 eV; transitions at exactly K originating from the bulk metal region have weaker contributions, with varying signs dependent on the layer thickness so that their summed contribution never outcompetes the near-K ones. This variation of the weak bulk contribution with respect to layer thickness is consistent with a previous theory result (*18*), where $\chi^{(2)}_{yyy}$ for the Al(111) surface was calculated to vary with Al thickness and eventually converge to 0.2 nm²/V.

In the decomposition analysis described above, we focused on so-called 2ω-terms. The other so-called interband and intraband ω-terms are ignored here because they cancel in a wide energy range near 1.5 eV. This cancellation is also found for many other solid-state systems (*35*).

Theory estimate of out-of-plane $\chi^{(2)}$

To demonstrate that the 2D metal systems supports a $\chi^{(2)}_{zzz}$ tensor element that is at least of the same magnitude as $\chi^{(2)}_{yyy}$, we calculated the static second harmonic optical response $\chi^{(2)}(\omega=0)$ of 2D-In following the first-principles finite-difference approach of Ref. (*36*). This model only captures the adiabatic response of the metal and excludes other factors such as interband transitions and plasmon resonances. $\chi^{(2)}_{zzz}$ at finite frequencies (at least below the plasma frequency of 2D-In) are expected to be at similar orders of magnitude (*37*). Following the Rudnick-Stern parametrization (*16*, *17*), $\chi^{(2)}_{zzz}$ is given by

$$\chi^{(2)}_{zzz} = a(\omega) \frac{e}{16\,\pi m \omega^2} [\varepsilon(\omega) - 1]$$

where $\varepsilon(\omega)$ is the dielectric function and $a(\omega)$, in its static limit, is a dimensionless parameter calculated by integrating over the polarization change $P_2$ at the second-order, following the second-order change in charge density $n_2$ in response to an external field (*36*, *38*, *39*)

$$a = \int_{-\infty}^{+\infty} dz\, P_2(z)$$

$P_2(z) = \int_{-\infty}^{z} dz'\, n_2(z')$

The first- and second-order perturbations in charge density can be obtained from finite-difference calculations ($n_+$, $n_0$, and $n_-$ from calculations with fields at $+E$, $0$, and $-E$ respectively) by $n_1 = (n_+ - n_-)/2$ and $n_2 = (n_+ + n_- - 2n_0)$. Benchmark calculations on surfaces of simple metals such as Al and Na shows that calculated $\chi^{(2)}$ compares favorably with experiments. The calculated second order polarizations and charge densities are shown in Figure S9(a,b). Vertical lines mark the height of the last layer of metal atoms extending from left to right; To its right, a charge density tail extends into the vacuum for both cases and to its left, the induced density decays into the bulk with Friedel oscillations. From these results, the *a* parameter for a Al(111) surface is estimated to be ~28, yielding a $\chi^{(2)}_{zzz}$ of ~1 nm²/V, consistent with the experimentally determined value of 1.0 nm²/V.

The same method is then applied to the In/SiC system, with results shown in Figure S9(c) along with the calculations cell aligned along the *z* direction. Within the bulk part of the semiconducting 6H-SiC substrate, the induced second-order polarization $P_2(z)$ averages to 0.13 a.u., equivalent to a bulk $\chi^{(2)}_{zzz}$ (in units of length/potential) of 25 pm/V, consistent with previous calculations at the density functional theory level (*40*) giving 11–28 pm/V. We remark on two distinctions between the semiconducting SiC and metallic In region. First, oscillations in $P_2(z)$ in SiC follows the positions of SiC units since this is a nonlinear response from a *semiconductor*, in contrast to Friedel oscillations in $P_2(z)$ in the metallic regions that follow the Fermi wavevector. Second, the similar magnitudes of the polarization $P_2$ in the two regions despite the drastically different orders of magnitude for $\chi^{(2)}_{zzz}$ (order of pm/V for SiC and nm/V for the metal) is because $\chi^{(2)}$ is defined with respect to the electric field *inside* the material; in the metal an electric field perpendicular to the surface is strongly screened.

In the bilayer indium region (green vertical lines), moderate contributions to $P_2(z)$ is observed *within* the metal layers, different from the rapidly decaying behavior in the simple metals discussed above. We attribute this to the bonding character changing over a longer distance in In/SiC (from covalent to metal) than in surfaces of simple metals. The *a* parameter for 2D-In is estimated to be ~18, yielding a $\chi^{(2)}_{zzz}$ of ~1 nm²/V.

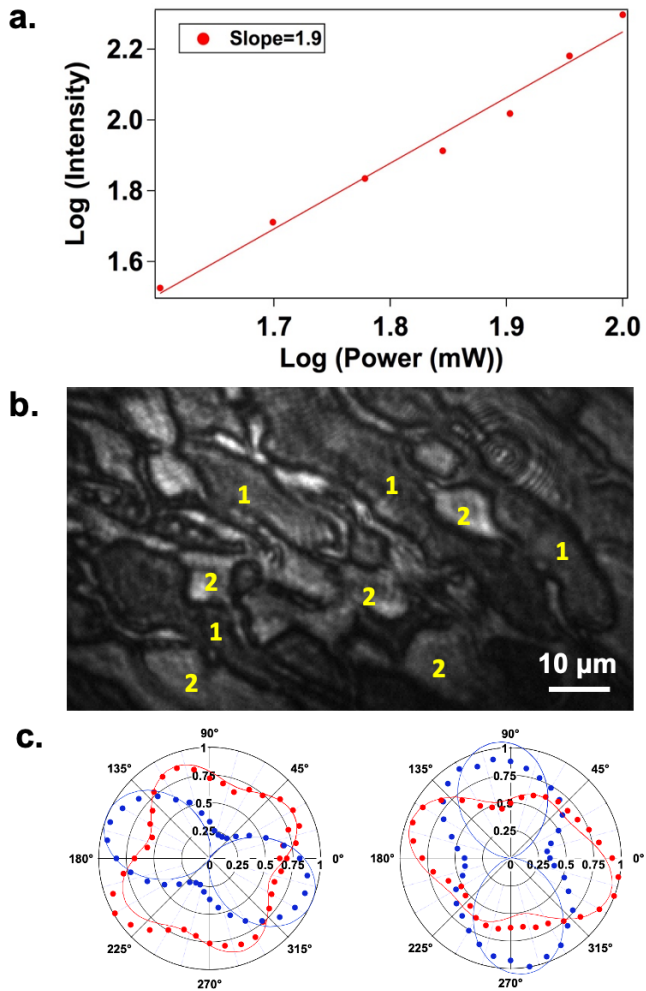

**Fig. S1.**

(a) Integrated intensity of the peak in the nonlinear spectrum of 2D-Ga (Figure 1b) at 3.1 eV, demonstrating a quadratic dependence on the incident power consistent with second harmonic generation. (b) SHG-detected image of 2D-Ga, with type 1 and type 2 regions labelled. (c) Excitation (red) and emission (blue) polar plots for type 1 (left) and type 2 (right) regions of 2D-Ga, collected as described in the text.

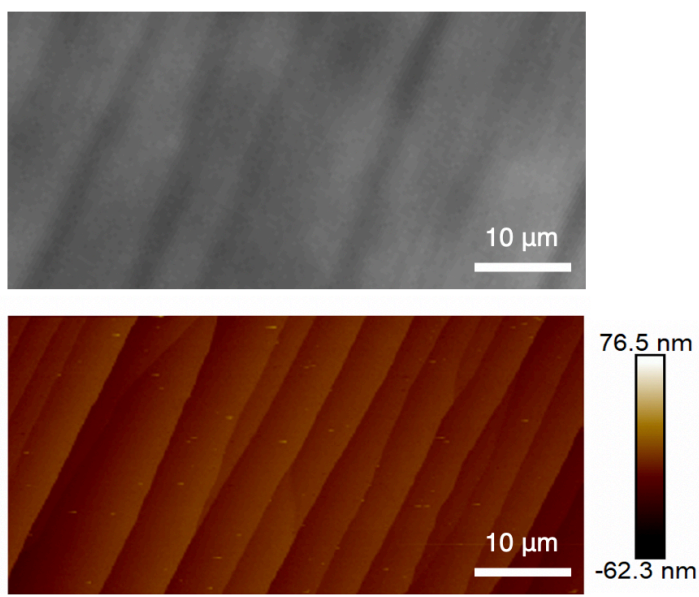

**Fig. S2.**
Comparison of optical (top) and atomic force microscopy (AFM) (bottom) images obtained from the 2D Indium sample depicted in Figure 3A. The images indicate that uniform terraces, persisting for several microns, are separated by step edges. Statistical analysis yields terrace width of 3 ± 1 μm (AFM) and 8 ± 2 μm.

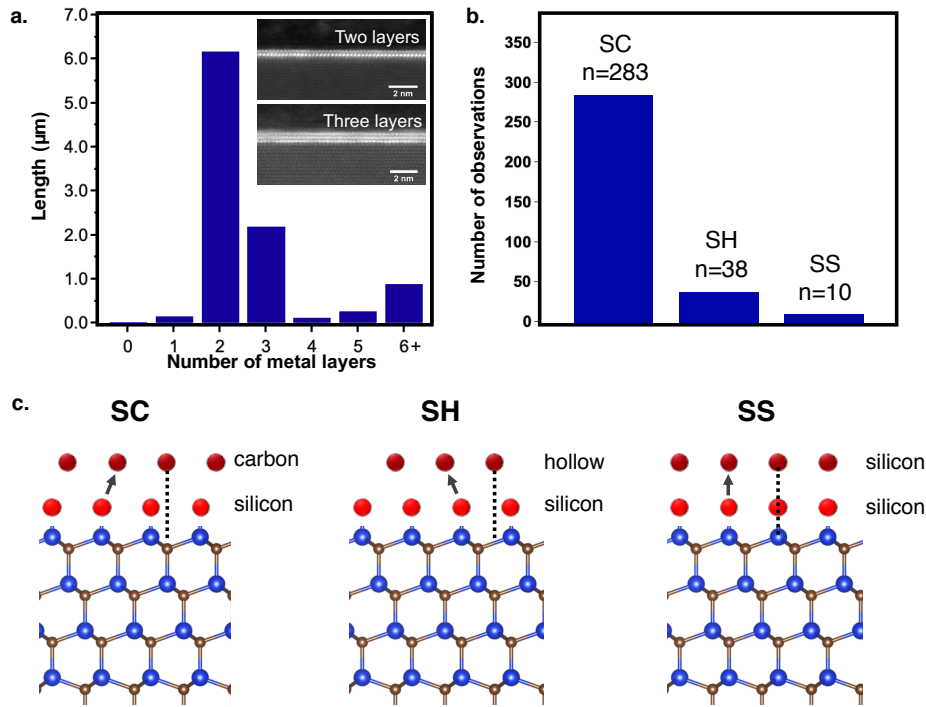

**Fig. S3**

(a) Thickness histogram of a 2D-In cross-section. Analysis of STEM images of 2D-In cross-section (9.8 μm total lateral distance) indicates that the 2D metal is predominately two and three atomic layers thick. Inset figures are STEM-HAADF images shows 2 and 3 layers of Indium.

(b) Analysis of several STEM cross-section images was used to approximate the prominence of three potential stacking types (c) for two layers of metal atoms with respect to the SiC substrate (silicon-carbon, silicon-hollow, and silicon-silicon). SC stacking is observed over 85% of the time, in agreement with theoretical predictions.

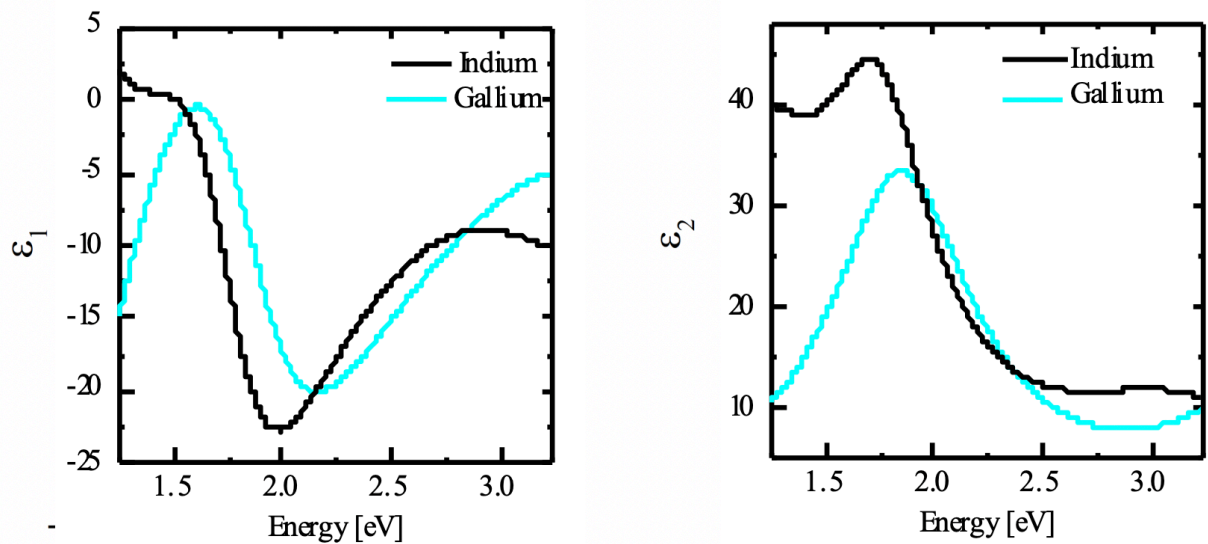

**Fig. S4**

Imaginary and real components of the dielectric function of 2D-In and 2D-Ga PMets obtained from spectroscopic ellipsometry.

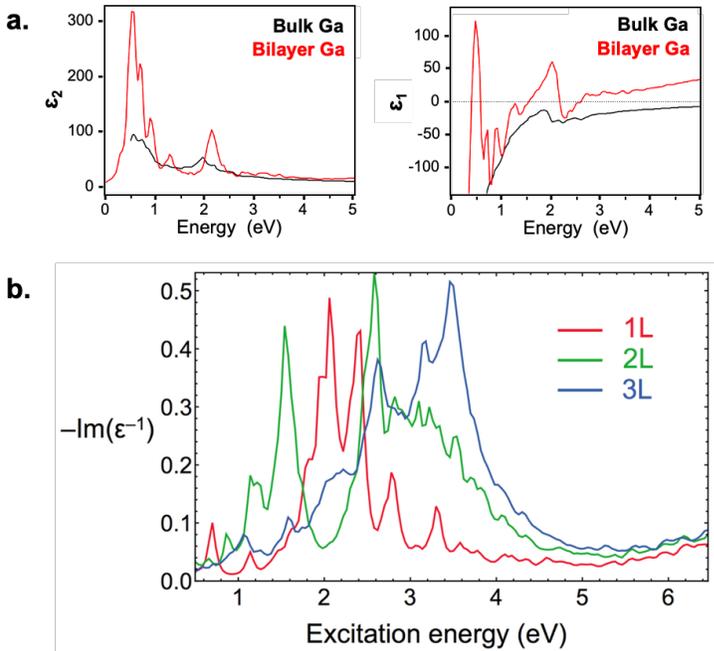

**Fig. S5**

(a) DFT- predicted real ($\varepsilon_1$) and imaginary ($\varepsilon_2$) components of the dielectric functions of bilayer and bulk gallium. (b) DFT-predicted plasmon resonances for 1, 2, and 3 layer gallium.

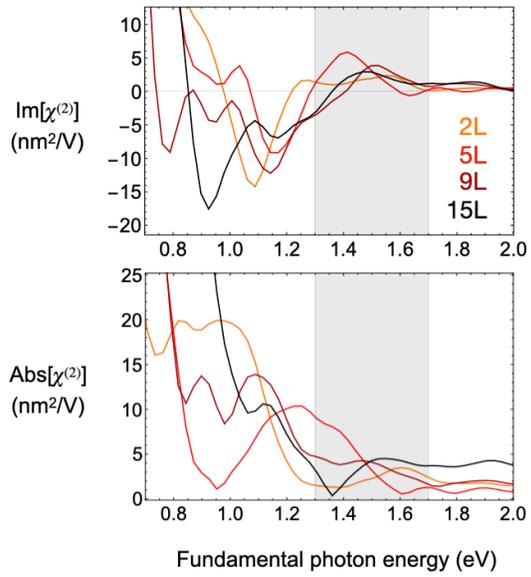

**Fig. S6**

Imaginary part and absolute value of $\chi^{(2)}_{yyy}$ calculated for 2,5,9 and 15 layers of indium on SiC, at different fundamental photon frequencies. The gray box covers an energy range of 1.5±0.2 eV.

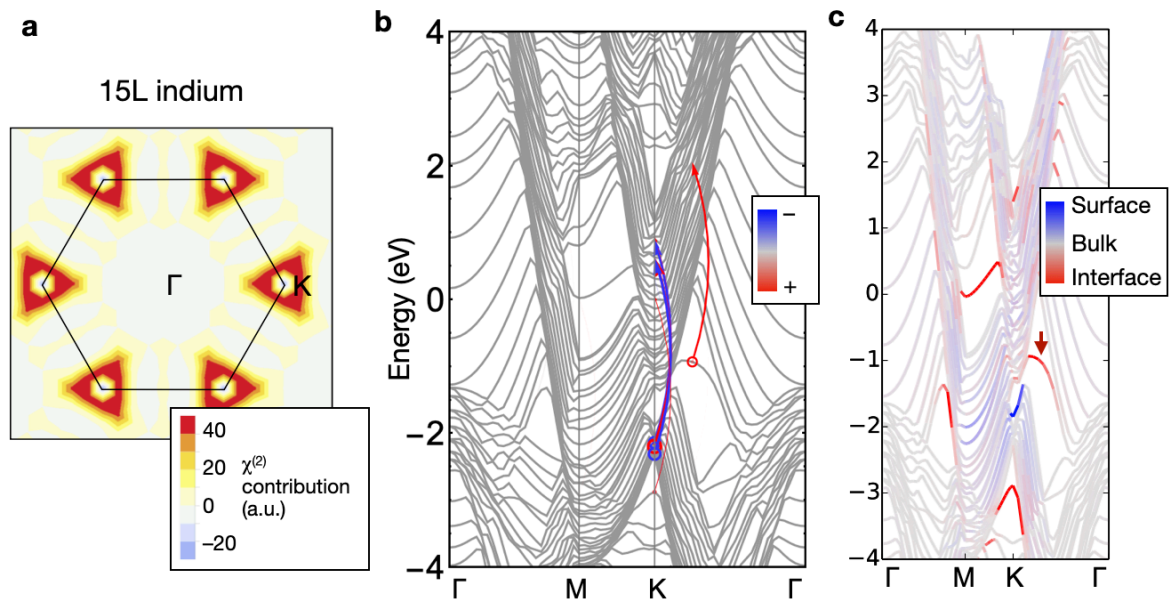

**Fig. S7**

**a,** Reciprocal space decomposition of $\chi^{(2)}$ for 15L-In/SiC within $\omega=1.5\pm0.2$eV. Dominant contributions appear close to K. **b,** Further decomposition by band contribution. Only the dominant transitions at $2\omega$ are visible. Red and blue indicate contributions of opposite signs. **c,** Same band structure but with red and blue colors indicating projection onto atomic orbitals of In atoms at the metal/SiC interface and at the surface.

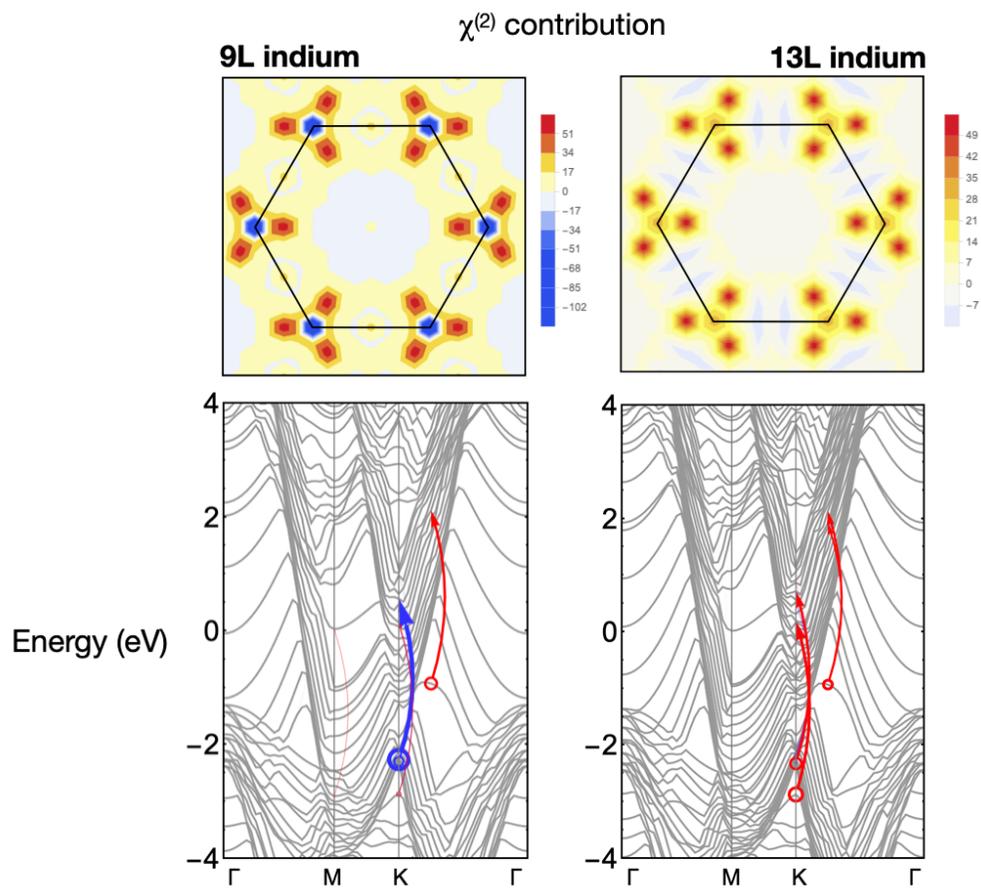

**Fig. S8**

Same as Fig. S7 but for 13L- and 9L-In/SiC

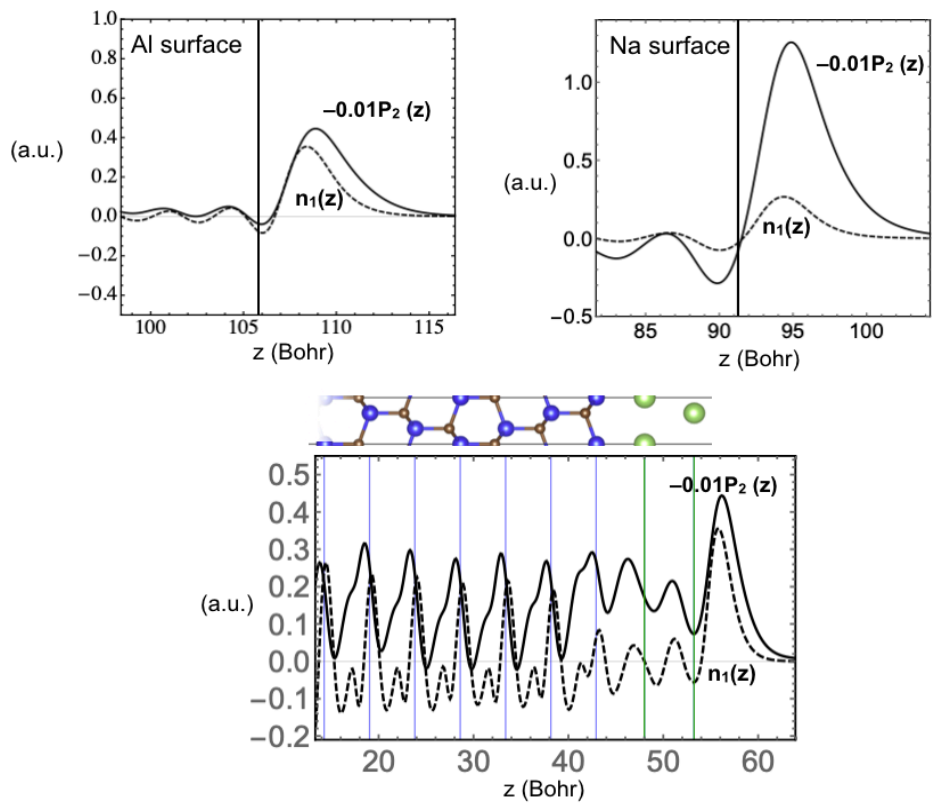

**Fig. S9**
(a,b) First-order response in charge density and second-order response in electronic polarization for the surfaces of two simple metals, Al(111) and Na(0001). Metal extends from left to right until the last atomic layer marked by the vertical line. (c) Same calculation for bilayer In/6H-SiC. Positions of Si and In atoms are respectively aligned with the plot using blue and green vertical lines.

**Table S1.**

Relative values for the elements of the $\chi^{(2)}$ tensor and $\theta$, the orientation of the crystal lattice, obtained from fitting the polarization patterns shown in Figure 3d,e. The relative values for the $\chi^{(2)}$ for a crystal with perfect $C_{3v}$ symmetry is shown for comparison.

**Type 1**

$$\begin{pmatrix} \chi^{(2)}_{xxx} & \chi^{(2)}_{xyy} & \chi^{(2)}_{xyx} \\ \chi^{(2)}_{yxx} & \chi^{(2)}_{yyy} & \chi^{(2)}_{yxy} \end{pmatrix} / \chi^{(2)}_{yyy} = \begin{pmatrix} -0.03\pm0.01 & -0.03\pm0.01 & -1.28\pm0.01 \\ -1.28\pm0.00 & 1 & -0.01\pm0.01 \end{pmatrix} \approx \begin{pmatrix} 0 & 0 & -1 \\ -1 & 1 & 0 \end{pmatrix}$$

$C_{3v}$

$\theta = 60 \pm 1°$

**Type 2**

$$\begin{pmatrix} \chi^{(2)}_{xxx} & \chi^{(2)}_{xyy} & \chi^{(2)}_{xyx} \\ \chi^{(2)}_{yxx} & \chi^{(2)}_{yyy} & \chi^{(2)}_{yxy} \end{pmatrix} / \chi^{(2)}_{yyy} = \begin{pmatrix} 0.00\pm0.01 & 0.02\pm0.01 & 0.66\pm0.01 \\ 0.68\pm0.01 & -1 & -0.01\pm0.02 \end{pmatrix} \approx \begin{pmatrix} 0 & 0 & 1 \\ 1 & -1 & 0 \end{pmatrix}$$

$C_{3v}$

$\theta = 66 \pm 4°$

**Movie S1.**

Excitation polarization-dependent SHG in 2D-In. Series of SHG-detected images of 2D-In obtained by changing , the polarization of the 800nm fundamental, as described in the text.

**Movie S2.**

Emission polarization-dependent SHG in 2D-In. Series of SHG-detected images of 2D-In obtained by changing , the polarization of the detected second harmonic signal, as described in the text.